# Large-area Waveguide Energy Harvesting Based on Fully-Polarized Elastic Topological Metamaterials

**Hanbang Deng[#], Bowei Wu[#], Tingfeng Ma[*], Teng Wang, Kun Hong**

Zhejiang-Italy Joint Lab for Smart Materials and Advanced Structures, School of Mechanics and Engineering Science, Ningbo University, Ningbo 315211, China;

## Abstract

To address the challenge that elastic-wave energy can only transmit through narrow-path waveguides in topological metamaterials, the proposal of large-area waveguides effectively breaks through this technical bottleneck. Nevertheless, elastic waves are vector waves with complex multi-component transmission characteristics. Realizing the cooperative transmission of in-plane and out-of-plane fully-polarized components poses substantial challenges for elastic-wave energy transmission and trapping applications. To tackle the above-mentioned problems, this paper proposes a fully-polarized elastic topological heterostructure based on the quantum valley Hall effect. First, symmetric unit-cell structures are designed to obtain unit cells with distinct topological properties for in-plane and out-of-plane modes, and multiple types of supercell structures are fabricated to realize the simultaneous transmission of fully-polarized elastic-wave energy for both in-plane and out-of-plane components. Furthermore, a gradient valley-locked structure is designed using large-area waveguide states to constrain and converge the transmitted energy, and its energy-harvesting performance is analyzed. The results demonstrate that the proposed structure enables coupled transmission of fully-polarized elastic-wave components. Moreover, the energy-harvesting capability of the gradient valley-locked phononic crystal plate is approximately 4.79 times that of conventional single-component transmission structures, which greatly improves the efficiency and transmission stability of acoustic energy harvesting. This work provides new insights for the engineering application of topological metamaterials in the field of energy harvesting.

[#] These authors contributed equally to this work.

[*] Authors to whom correspondence should be addressed: matingfeng@nbu.edu.cn

## I. Introduction

Efficient harvesting and stable transmission of acoustic wave energy have become one of the key technologies driving the development of low-power electronic devices. Typical applications include vibration-energy-harvesting power supply systems for equipment condition-monitoring sensors in industry, as well as acoustic energy harvesting and transport for powering implanted biomedical acoustic devices and acoustic-based portable diagnostic-treatment equipment in medicine. To date, numerous acoustic-energy-harvesting devices have been developed by researchers based on diverse mechanisms. Among them, piezoelectric acoustic devices have attracted considerable attention owing to the high energy-conversion efficiency and stability of piezoelectric materials [1–3]. Nevertheless, existing devices of this type generally suffer from several technical limitations, such as low energy-capture efficiency and severe losses caused by environmental disturbances and structural defects during energy transmission. These issues severely restrict the further development of this technology [4, 5].

To address the above-mentioned technical bottlenecks, elastic metamaterials and phononic crystals have been widely introduced into this field. By precisely tuning the structural geometric parameters and spatial arrangement, elastic metamaterials can realize localized confinement, directional focusing and efficient transmission of acoustic waves, thereby strengthening the performance of piezoelectric transducers and greatly improving the energy-harvesting efficiency [6, 7]. In addition, piezoelectric materials can be integrated into defect sites by introducing defects in phononic crystals for further energy-harvesting applications. However, plenty of challenges remain to be resolved in existing technical schemes. The defect modes of elastic metamaterials are susceptible to fabrication-induced structural imperfections [8–10]; the energy-capturing and transmission performances are vulnerable to disturbances from structural irregularities, leading to insufficient stability and reliability, which makes them difficult to satisfy the application requirements of complex industrial environments and extreme working conditions [11].

In recent years, the discovery of phononic topological insulators originating from condensed-matter physics has opened up new avenues for metamaterials to achieve efficient energy harvesting [12–15]. Different from the wave-modulation mechanism of conventional acoustic metamaterials, phononic topological insulators construct structural interfaces with distinct topological invariants to excite topologically protected edge states for topological transmission [16–18]. This special transmission mode does not rely on the microscopic geometric details of structures and is only determined by the global topological properties of the system, endowing devices with unique topological protection characteristics. Consequently, even when encountering geometric perturbations, structural defects or even local damage during energy transmission, significant scattering or loss will not occur, and energy can still propagate stably and efficiently along pre-defined paths [19–22]. Nevertheless, most topological edge states confine energy within an extremely narrow interface, which evidently limits high-capacity energy transmission and restricts their application scope [23,24].

Recently, topological heterostructures in metamaterials have provided a promising platform for constructing large-volume phononic topological devices [25,26]. The concept of topological heterostructures was first proposed in studies on valley-Hall phononic topological insulators. Its core design idea is to insert a heterostructure layer between two structural domains with different topological properties, so as to construct a unique large-area waveguide state that enables stable large-volume energy transmission [27–29]. On this basis, a series of novel topological devices featuring large-area waveguide states have been successfully realized and verified in various research systems including acoustics, photonics and elastic metamaterials [30–39]. Nevertheless, most existing research works with large-area waveguide structures can only transmit the out-of-plane component of elastic waves during propagation, while the transmission of in-plane components is neglected [40–42]. Owing to the complex modal coupling of elastic waves, energy transmission for both in-plane and out-of-plane components cannot be ignored. It still remains a challenge to realize energy harvesting and efficient transmission for fully-polarized in-plane and out-of-plane components of vector elastic waves [43–45].

To solve the above-mentioned issues, a special phononic topological heterostructure based on the quantum valley Hall effect is proposed and designed in this work. It can not only maintain the stable transmission of large-volume energy under large-area waveguides, but also simultaneously transmit the energy of both in-plane and out-of-plane components of vector elastic waves, so as to harvest more energy. In addition, the gradient valley-locking feature within the phononic crystal can constrain and converge energy during propagation, enabling energy to arrive at the receiver in a more concentrated manner and improving the energy-harvesting efficiency.

## II. Structural Design

As elastic waves are vector waves, making full use of their multi-component propagation characteristics of in-plane and out-of-plane components is of great significance for application exploration as well as efficient signal transmission and trapping. In general, the propagation of elastic waves in three-dimensional media satisfies the following equation:

$$\rho(\boldsymbol{r})\ddot{\mathbf{u}} = \nabla\{[\lambda(\boldsymbol{r}) + 2\mu(\boldsymbol{r})]\nabla\cdot\mathbf{u}\} - \nabla\cdot[\mu(\mathbf{r})\nabla\cdot\mathbf{u}] \tag{1}$$

where $\mathbf{u} = (u_x, u_y, u_z)$ denotes the displacement vector, $\lambda = E\nu/[(1+\nu)(1-2\nu)]$ and $\mu = E/2(1+\nu)$ represent the relationships between Lamé constants, Young's modulus E and Poisson's ratio $\nu$. In conventional studies, most investigations only consider the transmission of the out-of-plane component of $\mathbf{u}$, while the in-plane component is rarely explored.

Herein, a special $C_{6V}$ structure is adopted (as shown in Figure 1). Through elaborate structural design, the symmetry of the unit-cell structure is improved, and the double Dirac-cone degeneracy of out-of-plane and in-plane modes in dispersion curves is realized, which enables the co-transmission of in-plane and out-of-plane components of vector elastic waves. The phononic crystal based on the $C_{6V}$ structure is fabricated from aluminum, whose elastic modulus E is 68.9 GPa, density $\rho$ is 2700 kg/m$^3$, and Poisson's ratio $\nu$ is 0.33. The main geometric parameters of the unit-cell structure in the phononic crystal plate are illustrated in Figure 1, and the dispersion curves of the unit cell are mainly determined by geometric variables $b_1$, $b_2$, $m_1, m_2, d_1$ and $d_2$. Based on the unit-cell structure shown in Figure 1, parameters $(c_1, f_1, c_2, f_2)$ and $(e_1, e_2)$ are introduced to tune the above-mentioned geometric parameters such as $b_1$ and $b_2$, so as to adjust the symmetry of the unit-cell structure. Multiple unit-cell structures with distinct topological features are designed for efficient topological transmission. The geometric variables can be expressed as follows:

$$b_1 = (1 + c_1)b_0, -1 < c_1 < 1 \tag{2}$$

$$b_2 = (1 + f_1)b_1, -1 < f_1 < 1 \tag{3}$$

$$m_1 = (1 + c_2)b_0, -1 < c_2 < 1 \tag{4}$$

$$m_2 = (1 + f_2)m_1, -1 < f_2 < 1 \tag{5}$$

$$d_1 = e_1\left(\frac{l-b_1}{2} - \frac{t}{\sqrt{3}}\right), 1 \ll e_1 \ll 1 \tag{6}$$

$$d_2 = e_2\left(\frac{l-b_2}{2} - \frac{t}{\sqrt{3}}\right), 1 \ll e_2 \ll 1 \tag{7}$$

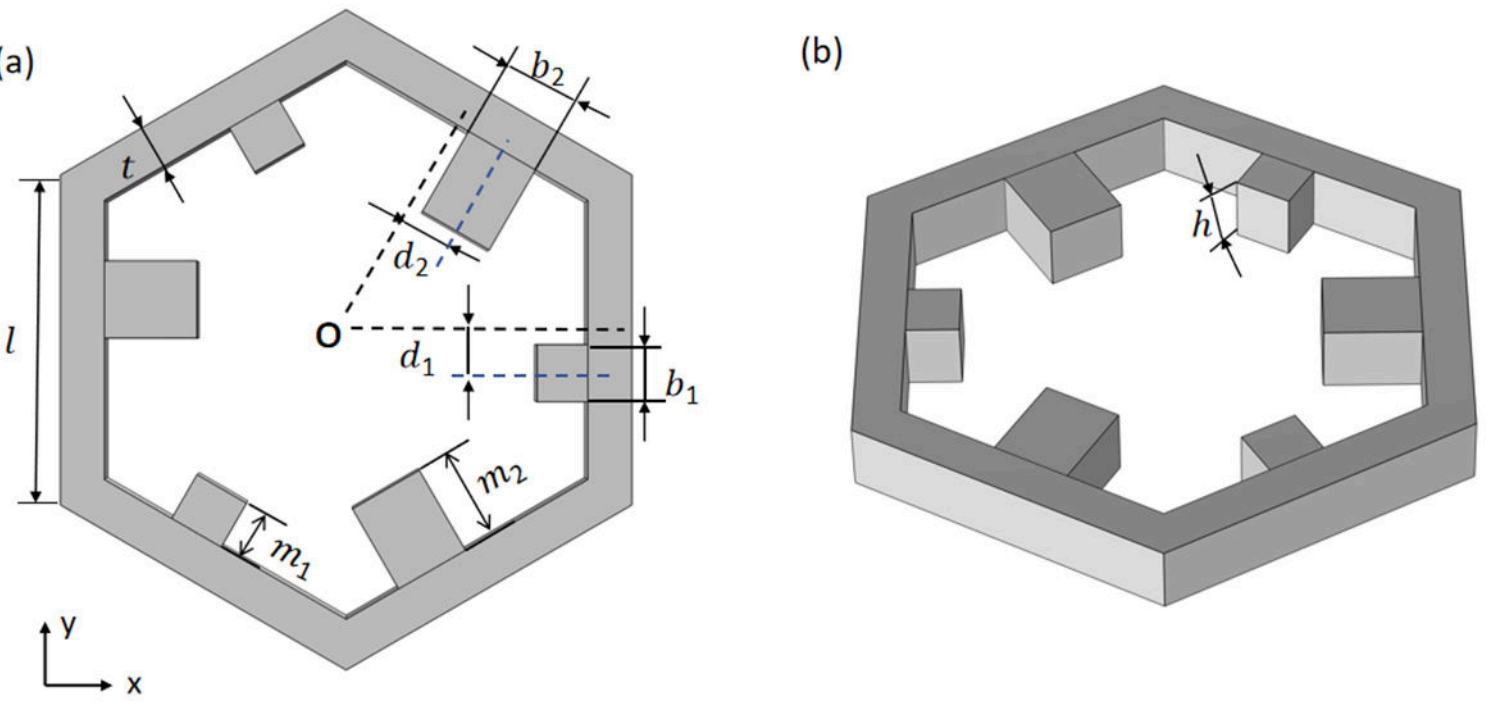

**FIG.1.** Schematic diagram of the $C_{6V}$ structure and geometric parameters, where l=15mm,a=$\sqrt{3}$l, h=3mm, t=2mm, $b_0$=2mm

Subsequently, unit-cell models with different geometric parameters are constructed using the COMSOL Multiphysics simulation software, and their dispersion characteristics are calculated and analyzed. Unit-cell dispersion curves under multiple states of Dirac degeneracy and degeneracy-lifting are obtained, as shown in Figure 2. First, calculations are performed to determine that Unit-Cell Type-A is obtained when the parameters$(c_1, f_1, c_2, f_2)$and $(e_1, e_2)$ are set to (0.3, 0.35, 0.15, 0.83) and (0,0), respectively. For this unit cell, the dispersion curves of out-of-plane and in-plane modes exhibit simultaneous degeneracy near 61.5 kHz at the K-point of the wave vector, forming a double Dirac cone, as depicted in Figure 2(a). The red dots denote out-of-plane modes, while the blue dots represent in-plane modes. The quantification of in-plane and out-of-plane motion modes for this unit-cell structure is mainly determined by the polarization factor $w_z$, which is expressed as:

$$w_z = \frac{\int_{V_c} |z|^2 dV}{\int_{V_c} (|x|^2 + |y|^2 + |z|^2) dV} \quad (8)$$

where x, y and z correspond to the components of elastic waves along the x, y and z directions, and $V_c$ is the volume of the unit cell. When $w_z$ is close to 1, the out-of-plane component of the elastic wave dominates. In contrast, when $w_z$ approaches 0, the in-plane component becomes dominant.

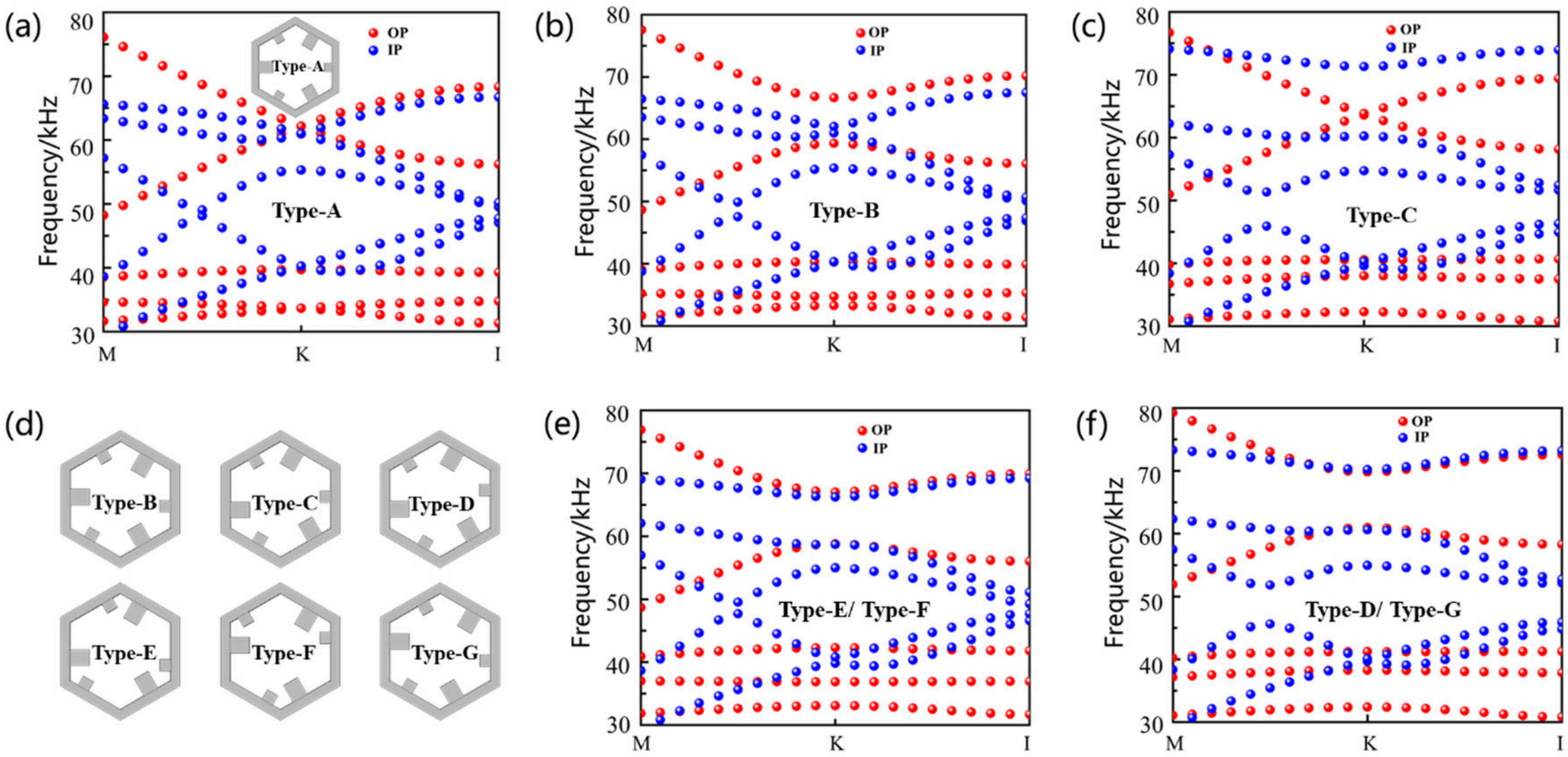

**FIG. 2.** (a) Model diagram and dispersion curves of Unit-Cell A. (b) Dispersion curves of Unit-Cell B. (c) Dispersion curves of Unit-Cell C. (d) Model diagrams of Unit-Cells B, C, D, E, F and G. (e) Dispersion curves of Unit-Cell E or F. (f) Dispersion curves of Unit-Cell D or G, where red dots denote out-of-plane modes and blue dots denote in-plane modes.

Subsequently, starting from the unit cell with a double Dirac cone, multiple unit-cell structures with topologically nontrivial states are constructed by keeping $(c_1, f_1, c_2, f_2)$ fixed at (0.3, 0.35, 0.15, 0.83) and tuning the parameter combinations of (e1,e2). It is observed that Unit-Cell B is obtained when (e1,e2) equals (0.3, 0.1), whose dispersion curves are presented in Figure 2(b). Only the dispersion curve of the in-plane mode exhibits single-point degeneracy, while the degeneracy point of the out-of-plane mode splits and opens a band gap. When(e1,e2) is set to (-0.1, -0.5), Unit-Cell C is achieved, as shown in Figure 2(c). A single-point degeneracy exists only for the out-of-plane mode, whereas the degeneracy point of the in-plane mode splits to form a band gap. Based on the dispersion characteristics and geometric parameters of the above unit cells, unit-cell dispersion curves under different (e1,e2) parameter combinations are calculated, so as to obtain unit cells with simultaneous lifting of the double Dirac-cone degeneracy. Various unit-cell models are displayed in Figure 2(d).

Take the tuned Unit-Cell Type-D (-0.4, -0.47)and Unit-Cell Type-E (0.6, -0.3) as examples, whose models are illustrated in Figure 2(d). Calculation and analysis of their dispersion curves reveal that the Dirac-degenerate states of both out-of-plane and in-plane modes are lifted simultaneously, generating complete band gaps, as shown in Figure 2(f) and Figure 2(e), respectively. Unit-Cell Type-F (-0.6, 0.3) and Unit-Cell Type-G (0.4, 0.47)exhibit similar complete-band-gap behaviors to Unit-Cell Type-E and Unit-Cell Type-D, respectively. Nevertheless, although these four unit cells share similar dispersion-curve features, pairwise differences in topological properties exist among them.

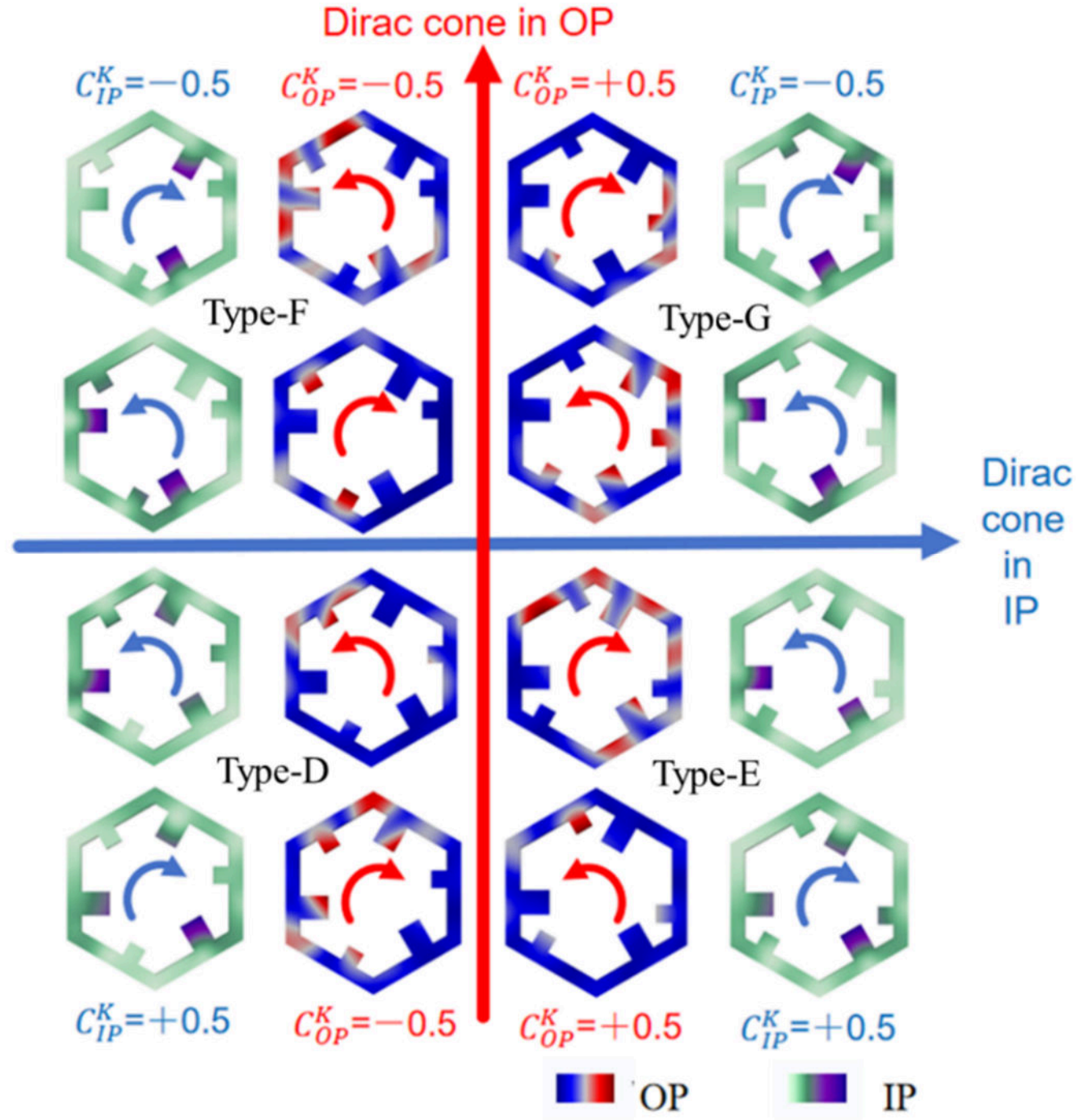

**FIG. 3.** In-plane and out-of-plane displacement modes of different unit cells.

To analyze the topological inversion phenomenon of different unit cells, the in-plane and out-of-plane motion modes of various unit cells at the K-point of the wave vector are presented in Figure 3. It can be observed that Unit-Cell Type-D and Type-E possess opposite out-of-plane motion directions but identical in-plane motion directions, which verifies the topological inversion of the out-of-plane mode between Unit-Cell Type-D and Type-E. Similarly, topological inversion of the in-plane mode occurs between Unit-Cell Type-D and Type-F; for Unit-Cell Type-D and Type-G, topological inversion takes place for both out-of-plane and in-plane modes. Such differences in out-of-plane and in-plane motion modes among these unit-cell structures with distinct topological properties lay the foundation for constructing topological edge states and realizing topological transmission for different elastic-wave modes.

In addition, when Unit-Cell Type-A with a double Dirac cone is placed at the topological interfaces formed pairwise by the above four unit cells, large-area pseudo-spin-momentum-locked helical waveguide states can emerge, which also provides a basis for high-efficiency topological transmission and energy harvesting.

## III. Topological Edge States of Elastic Waves for Different Modes

Based on the aforementioned phononic-crystal structures, multiple supercell structures composed of two or three unit cells with distinct topological properties are constructed, and their respective displacement-field distributions and dispersion curves are calculated. Accordingly, the topological edge states generated among these phononic-crystal structures as well as the transmission behaviors of in-plane and out-of-plane elastic-wave components within the topological boundaries are analyzed.

As shown in Figure 4(b), Supercell-1 composed of Unit-Cell Type-D and Unit-Cell Type-E is first constructed, and its displacement-field distributions and dispersion curves are calculated. It can be observed from the results that only one out-of-plane topological edge state (marked in red) exists within the band-gap region of Supercell-1 ranging from 66 kHz to 72 kHz. This indicates that the topological interface formed by Unit-Cell Type-D and Type-E only permits the transmission of the out-of-plane component of vector elastic waves, whereas the in-plane elastic-wave component cannot propagate. This transmission behavior is also adopted in numerous existing studies.

Subsequently, Supercell-2 consisting of Unit-Cell Type-D and Unit-Cell Type-F is constructed, and its displacement-field distributions and dispersion curves are calculated, as shown in Figure 4(c). It can be seen from the figure that only one in-plane topological edge state (marked in blue) appears within the band-gap range of Supercell-2 from 66 kHz to 72 kHz. This demonstrates that the topological interface formed by Unit-Cell Type-D and Unit-Cell Type-F only allows the transmission of the in-plane component of vector elastic waves, while the out-of-plane elastic-wave component cannot propagate.

On the basis of the differences in motion modes of the above-mentioned two topological interfaces, we further explore the scenario where topological edge states for both in-plane and out-of-plane elastic-wave modes coexist at the topological boundary. Supercell-3 composed of Unit-Cell Type-D and Unit-Cell Type-G is adopted, and its displacement-field distributions and dispersion curves are calculated, as shown in Figure 4(c). Two distinct topological edge states emerge within the band-gap range of 66 kHz to 72 kHz in the dispersion plot, where the red curve corresponds to the out-of-plane mode and the blue curve corresponds to the in-plane mode. This reveals that both in-plane and out-of-plane components of vector elastic waves can be transmitted simultaneously at the topological interface formed by Unit-Cell Type-D and Unit-Cell Type-G, leading to more sufficient and efficient energy transmission compared with the former two topological transmission schemes.

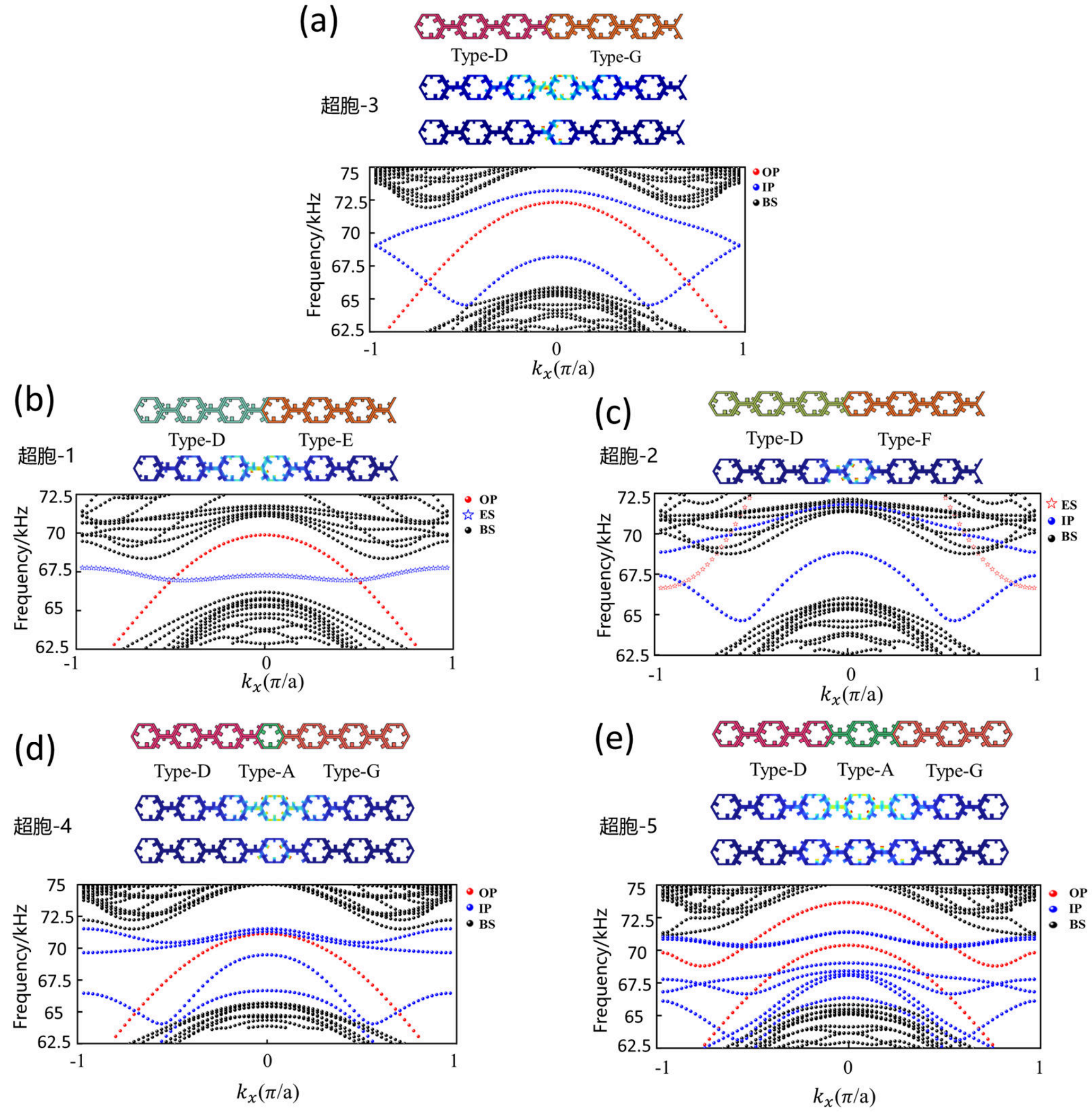

**FIG. 4.** (a) Schematic diagram, displacement-field distribution and dispersion curves of Supercell-3. (b)-(e) Schematic diagram, displacement-field distribution and dispersion curves of Supercell-1, 2, 3, 4, 5, respectively.

Accordingly, Unit-Cell Type-A with a double Dirac cone is introduced at the topological interface of Supercell-3. A supercell consisting of Unit-Cell Type-A, Type-D and Type-G is constructed, and its dispersion curves are calculated to investigate the characteristics of the topological edge states. Supercell-4 is first built by inserting one Unit-Cell Type-A at the interface between Unit-Cell Type-D and Type-G, and the corresponding dispersion curves are obtained, as shown in Figure 4(d). It can be observed that several distinct topological edge states still exist within the band-gap range from 66 kHz to 72 kHz, where the red curves correspond to out-of-plane modes and the blue curves correspond to in-plane modes. This indicates that when Unit-Cell Type-A featuring a double Dirac cone is introduced into the topological interface formed by Unit-Cell Type-D and Type-G, the topological edge states for both in-plane and out-of-plane modes can still be preserved, enabling the simultaneous transmission of in-plane and out-of-plane components of vector elastic waves.

To further explore the influence of the number of Unit-Cell Type-A on the in-plane and out-of-plane motion modes of topological edge states in this structure, dispersion curves of supercell structures under various configurations are further calculated. Herein, Supercell-5 is formed by inserting three Unit-Cell Type-A at the interface between Unit-Cell Type-D and Type-G. As illustrated in Figure 4(e), several distinct topological edge states can still be observed within the band-gap range of 66 kHz to 72 kHz, where red curves correspond to out-of-plane modes and blue curves correspond to in-plane modes. This demonstrates that when the number of inserted Unit-Cell Type-A with a double Dirac cone varies within a certain range at the topological interface formed by Unit-Cell Type-D and Type-G, the phononic-crystal structure can still sustain the topological edge states of both in-plane and out-of-plane modes and realize the simultaneous transmission of in-plane and out-of-plane components of vector elastic waves.

## IV. Transmission Characteristics of Topological Edge States for Elastic Waves and Gradient Valley-Locked Waveguides

### A. Energy Transmission Characteristics of Elastic Waves

Based on the differences among various topological-edge-state modes at the aforementioned topological interfaces, several simulation models of phononic-crystal plates are designed and constructed using COMSOL Multiphysics in this section on the basis of different supercells. The elastic-wave transmission characteristics under various phononic-crystal-plate configurations are calculated and analyzed. Meanwhile, physical prototypes are fabricated via CNC machining, and experiments are designed to verify the transmission behaviors of elastic waves in these phononic-crystal plates.

First, Supercell-1, which only supports out-of-plane topological edge states at its topological interface, is periodically arranged along the x-direction to form a $1\times10$ phononic-crystal plate-1, as shown in Figure 5(a). In COMSOL Multiphysics, an elastic-wave excitation source (marked by the five-point star in Figure 5(a)) is applied at the central position on the left side of the phononic-crystal plate, with the excitation frequency ranging from 66 kHz to 72 kHz. A receiver is placed at the central position on the right side (denoted by the square in Figure 5(a)). Low-reflection boundaries are imposed on the four peripheral boundaries of the phononic-crystal plate to suppress reflected waves, so that more accurate transmission characteristics of the plate can be obtained.

Similarly, Phononic-Crystal Plate-2 and Phononic-Crystal Plate-3 are constructed using Supercell-2 and Supercell-3, respectively, as depicted in Figure 5(b) and Figure 5(c). Under identical COMSOL settings, the displacement-field distributions of vector elastic-wave signals in these phononic-crystal plates are calculated correspondingly.

In addition, in practical experiments, the three-dimensional displacement fields of the above-mentioned phononic-crystal plates are measured using the LV-FSC500-3D scanner. The excitation source at the input terminal on the left side of the phononic-crystal plate is driven by PZT-5A piezoelectric ceramics, with the excitation frequency ranging from 66 kHz to 72 kHz. A receiver is placed at the center of the right side, and the weak-reflection boundaries are simulated by covering the edges with sound-absorbing layers.

By observing and analyzing the obtained displacement-field distributions from COMSOL simulations and experiments, it can be found that both sets of results demonstrate stable transmission of elastic-wave components along topological interfaces. For Phononic-Crystal Plate-1, only the out-of-plane component can propagate stably along the topological interface, as shown in Figure 5(a). For Phononic-Crystal Plate-2, the in-plane component can achieve stable topological transmission, as presented in Figure 5(b). For Phononic-Crystal Plate-3, both out-of-plane and in-plane components of elastic waves can propagate simultaneously along the topological interface, as illustrated in Figure 5(c). Comparisons between experimental and simulated results reveal certain discrepancies. This originates from the damping effect of the aluminum material, which induces energy attenuation during elastic-wave propagation, whereas material damping is not taken into account in the numerical simulations.

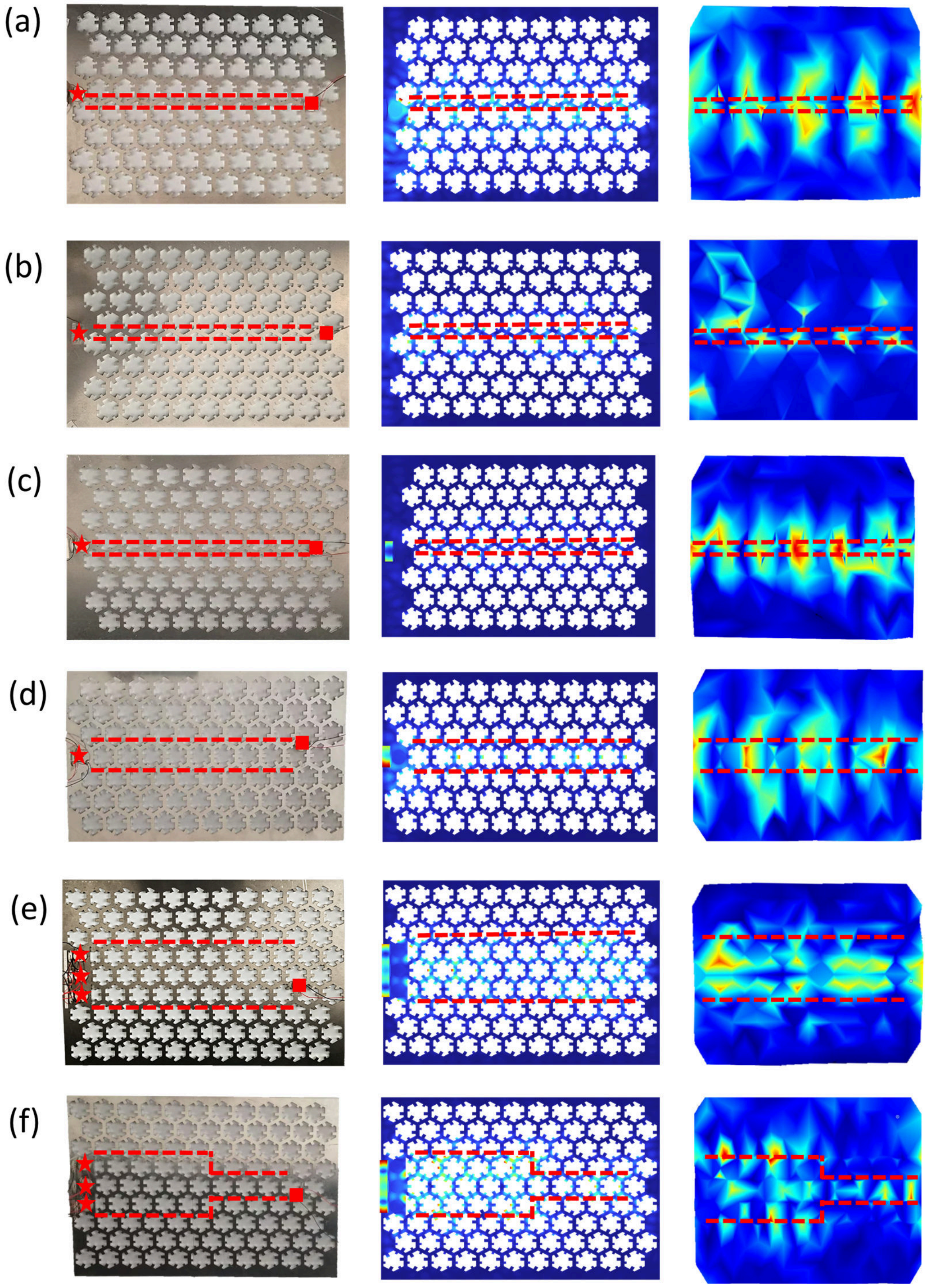

**FIG. 5.** (a) Physical prototype and displacement-field distribution of Phononic-Crystal Plate-1, where the five-point star denotes the excitation, the square denotes the response, and the region within the red dashed line represents the energy transmission channel. (b)-(f) Physical prototype and displacement-field distribution of Phononic-Crystal Plate-2, 3, 4, 5, 6, respectively.

However, by examining the simulated displacement-field results in Figure 5(a), Figure 5(b) and Figure 5(c), it can be found that the energy transmission channels of Phononic-Crystal Plate-1, Plate-2 and Plate-3 are confined merely to the interfaces between different unit cells. These channels are relatively narrow, leading to a limited volume for energy transmission. As mentioned earlier, inserting Unit-Cell Type-A with a double Dirac cone into the pairwise topological interfaces formed by the four unit cells can give rise to large-area waveguide states. Combined with the signal transmission behaviors of the three phononic-crystal plates, several Unit-Cell Type-A are inserted into Phononic-Crystal Plate-3 for structural optimization, so as to further increase the energy-transmission volume and improve the efficiency of signal transmission.

First, Supercell-4 containing one Unit-Cell Type-A (i.e., one valley-locked unit) is periodically arranged along the x-direction to form a 1×10 Phononic-Crystal Plate-4, as shown in Figure 5(d). Identical settings are adopted for both COMSOL simulations and physical experiments as those for Phononic-Crystal Plate-3, and the corresponding two sets of displacement-field distributions are obtained, as presented in Figure 5(d). Analysis of the results shows that both in-plane and out-of-plane components of elastic waves can still be stably transmitted. Meanwhile, comparing the displacement-field results in Figure 5(d) with those in Figure 5(c), it can be seen that under identical excitation, Phononic-Crystal Plate-4 introducing large-area waveguides exhibits an obvious expansion in energy-transmission volume compared with Phononic-Crystal Plate-3.

To make full use of the transmission characteristics of this structure, multiple Unit-Cell Type-A are inserted at the topological interface to further enlarge the waveguide area and increase the energy-transmission volume. As shown in Figure 5(e), Supercell-5 containing three Unit-Cell Type-A is periodically arranged along the x-direction to construct a 1×10 Phononic-Crystal Plate-5. Different from the configurations of the previous four phononic-crystal plates, three excitation sources are introduced here to fully exploit the large-area topological waveguide. They are placed on the left side following a certain rule (marked by five-point stars in the figure). Each excitation source applies the same prescribed displacement. The receivers on the right side as well as the peripheral boundaries remain consistent with those of other structures. Two sets of displacement-field distributions for this phononic-crystal plate are obtained via COMSOL simulations and physical experiments, as illustrated in Figure 5(e).

Result analysis reveals that both in-plane and out-of-plane components of elastic waves can still be stably transmitted. Moreover, by comparing the transmission paths of Phononic-Crystal Plate-5 and Phononic-Crystal Plate-4, it is found that the energy-transmission volume of Plate-5 is three times that of Plate-4, showing a multiple relationship corresponding to the number of inserted Unit-Cell Type-A.

Nevertheless, although the transmitted energy along the path can be greatly enhanced by enlarging the waveguide area, part of the energy will inevitably dissipate or deviate far from the receiver position owing to the large waveguide area and the positional limitation of the receiver. Accordingly, to achieve more concentrated energy and higher energy-transmission efficiency, we further innovate and optimize the structure of the phononic-crystal plate based on the above findings. Supercell-4 containing one Unit-Cell Type-A and Supercell-5 containing three Unit-Cell Type-A are periodically arranged along the x-direction in a 1×5×1×5 configuration to form Phononic-Crystal Plate-6 with gradient valley-locked waveguides, where the number of valley locks for energy transmission changes from three to one, as shown in Figure 5(f). Identical settings are maintained for both COMSOL simulations and physical experiments as those for Phononic-Crystal Plate-5, and the corresponding displacement-field distributions are obtained, as illustrated in Figure 5(f). It can be observed from the displacement-field results that both in-plane and out-of-plane elastic-wave energy can still be stably transmitted, and a gradient variation of waveguide area appears along the transmission path. It can be inferred that Phononic-Crystal Plate-6 with gradient valley-locked waveguides may realize the convergence and enrichment of energy inside the waveguide through such step-wise variation.

### B. Energy Harvesting of Elastic Waves

To investigate the practical performance of energy-transmission volume and efficiency for phononic-crystal plates with different configurations, the elastic-wave energy propagating in these plates is harvested and converted into electric energy for comparative analysis in this section. A circular PZT-5A piezoelectric ceramic with dimensions of φ15 × 1 mm is adopted for excitation at the input terminal on the left side of the phononic-crystal plate, where a sinusoidal signal is applied. Another circular

PZT-5A piezoelectric ceramic of φ15 × 1 mm is attached at the receiving terminal on the right side to serve as a piezoelectric energy harvester, as illustrated in Figure 5. Meanwhile, a pure aluminum plate with comparable overall dimensions is fabricated for comparison of electric output with each phononic-crystal plate. The excitation and receiving terminals of the pure aluminum plate are kept consistent with those of the aforementioned phononic-crystal plates.

First, excitation along the z-direction is applied to Phononic-Crystal Plate-1, which only allows the propagation of the out-of-plane component of elastic waves. The reference pure-aluminum plate for comparison is also excited by signals along the z-direction. After sweeping with sinusoidal excitation signals, we observe slight differences in the peak voltage at different frequencies within the range of 66 kHz-72 kHz from the open-circuit voltage spectrum recorded by the oscilloscope. This arises because the performance of the piezoelectric energy harvester at the receiver is strongly dependent on the strain field at the harvesting position. The strain field during elastic-wave propagation may vary at certain frequencies under different conditions, together with fabrication errors introduced in the manufacturing of physical prototypes, leading to fluctuations in the output voltage.

Accordingly, we select frequencies near their respective higher-voltage points for measurement. The relationships between output voltage and load resistance are measured for Phononic-Crystal Plate-1 at 69.06 kHz and the pure-aluminum plate at 69.05 kHz, as shown in Figure 6(a). It can be seen that the output voltages of both specimens increase with the increase of load resistance and tend to saturate when the resistance exceeds a certain threshold. The saturation voltage of Phononic-Crystal Plate-1 is 0.526 V, while that of the pure-aluminum plate is 0.229 V. According to Ohm’s law, the relationships between output power and load resistance for the two cases are calculated, as illustrated in Figure 7(a). The maximum output power of Phononic-Crystal Plate-1 at the corresponding frequency is 0.132 mW, whereas the maximum output power of the pure-aluminum plate is 0.019 mW. The comparison reveals that the voltage and power of Phononic-Crystal Plate-1 are approximately 2.3 times and 6.95 times those of the pure-aluminum plate, respectively.

Similarly, excitation along the x-direction is applied to Phononic-Crystal Plate-2, which only enables the transmission of in-plane components, as well as the reference pure-aluminum plate. Measurements are carried out at nearby frequencies corresponding to their relatively high-voltage points. The relationships between output voltage and load resistance are measured for Phononic-Crystal Plate-2 at 69.27 kHz and the pure-aluminum plate at 69.05 kHz, as shown in Figure 6(b). It can be observed that the output-voltage trends for both samples are consistent with those described above. The saturation voltage of Phononic-Crystal Plate-2 is 0.364 V, while that of the pure-aluminum plate is 0.182 V. According to Ohm's law, the relationships between output power and load resistance for the two cases are calculated, as illustrated in Figure 7(b). The maximum output power of Phononic-Crystal Plate-2 at the corresponding frequency is 0.076 mW, whereas the maximum output power of the pure-aluminum plate is 0.014 mW. The comparison shows that the voltage and power of Phononic-Crystal Plate-2 are approximately 2 times and 5.4 times those of the pure-aluminum plate, respectively.

Subsequently, excitations along both the z- and x-directions are simultaneously applied to Phononic-Crystal Plate-3, which supports the simultaneous transmission of both in-plane and out-of-plane elastic-wave components, and the reference pure-aluminum plate. Measurements are performed at nearby frequencies corresponding to their relatively high-voltage points. The relationships between output voltage and load resistance are measured for Phononic-Crystal Plate-3 at 69.12 kHz and the pure-aluminum plate at 66.95 kHz, as shown in Figure 6(c). The saturation voltage of Phononic-Crystal Plate-3 is 1.151 V, while that of the pure-aluminum plate is 0.302 V. According to Ohm's law, the relationships between output power and load resistance for the two cases are calculated, as illustrated in Figure 7(c). The maximum output power of Phononic-Crystal Plate-3 at the corresponding frequency reaches 0.7 mW, whereas the maximum output power of the pure-aluminum plate is 0.033 mW. The comparison indicates that the voltage and power of Phononic-Crystal Plate-3 are approximately 3.8 times and 21.21 times those of the pure-aluminum plate, respectively. Meanwhile, comparative analysis of the saturation voltages among Phononic-Crystal Plate-3 and the other two phononic-crystal structures reveals that the performance indices of Plate-3 are not only higher than those of the other two plates individually, but also exceed the sum of their respective values. This further verifies that Phononic-Crystal Plate-3 can not only propagate in-plane and out-of-plane elastic-wave components simultaneously, but also achieve certain coupling enhancement between them, thus improving the energy-transmission volume and efficiency.

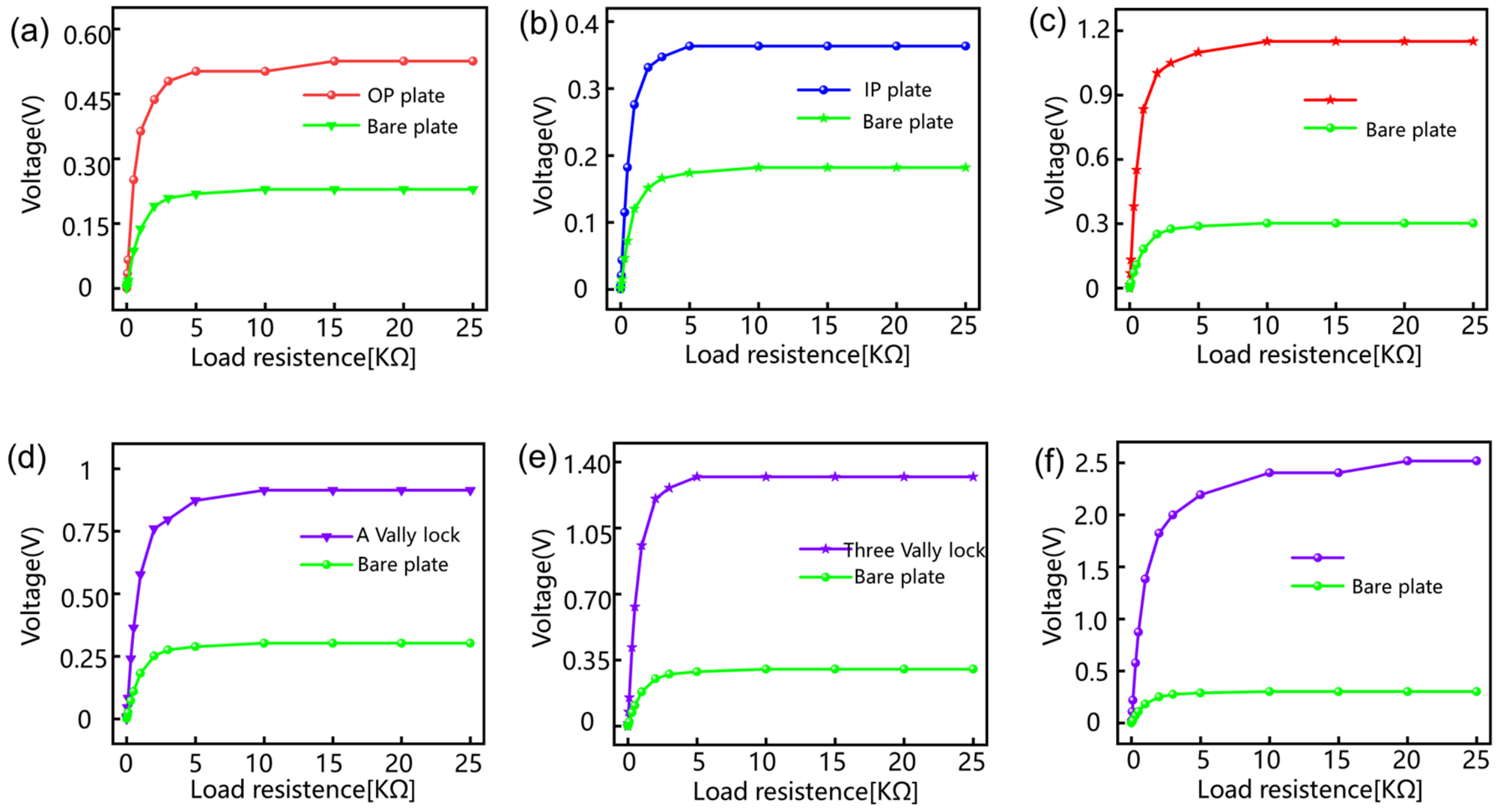

**FIG. 6.** (a) Relationship between output voltage and load resistance for Phononic-Crystal Plate-1. (b)-(f) Relationship between output voltage and load resistance for Phononic-Crystal Plate-2, 3, 4, 5, 6, respectively.

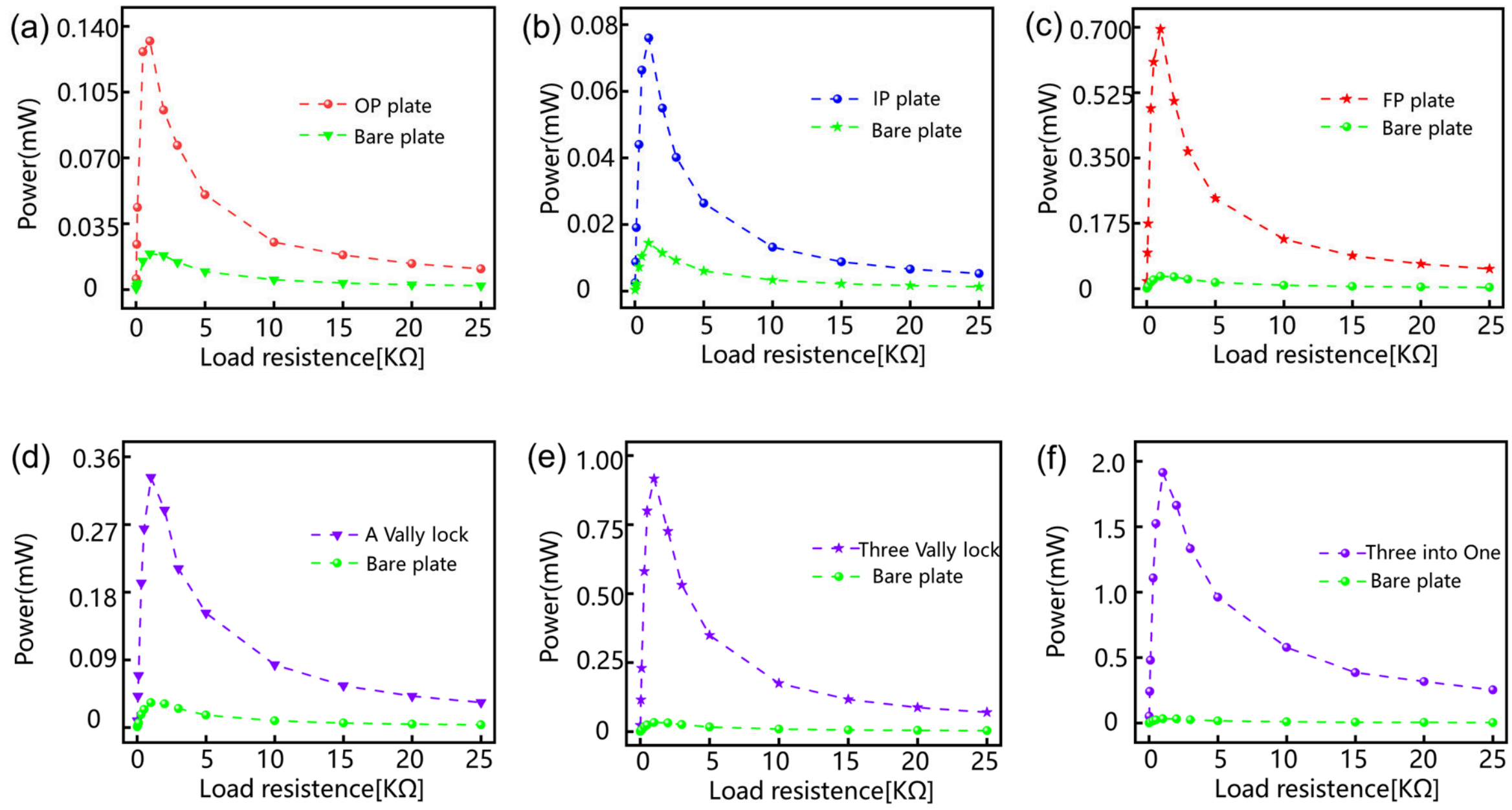

**FIG. 7.** (a) Relationship between output power and load resistance for Phononic-Crystal Plate-1. (b)-(f) Relationship between output power and load resistance for Phononic-Crystal Plate-2, 3, 4, 5, 6,respectively.

However, as demonstrated by the displacement-field results of various phononic-crystal plates in Figure 5, the energy-transmission channels of Phononic-Crystal Plate-1, Plate-2 and Plate-3 are confined to the interfaces between different unit cells. These channels are relatively narrow, resulting in a limited volume for energy transmission. When unit cells with a double Dirac cone are introduced into the narrow topological interfaces, large-area waveguide states can be generated, which enlarges the transmission volume of elastic-wave energy and improves the energy-harvesting efficiency.

To verify whether the phononic-crystal plates obtained by enlarging the waveguide area based on Phononic-Crystal Plate-3 (which enables full-polarization elastic-wave transmission) still maintain the capability for simultaneous transmission of in-plane and out-of-plane components, and to investigate whether energy inside the waveguide can be concentrated and converged to enhance the received energy at the receiver by introducing gradient variation into the large-area waveguide, the elastic-wave responses of Phononic-Crystal Plate-4, Plate-5 and Plate-6 are measured.

For Phononic-Crystal Plate-4, the excitation source at the input terminal, the receiver, and the peripheral boundaries are kept identical to those of Phononic-Crystal Plate-3. Excitations along both the x- and z-directions are applied simultaneously, and the receiver is placed at the center of the right side. Sampling is performed at frequencies corresponding to relatively high output voltages within the frequency range of 66 kHz-72 kHz. As shown in Figure 6(d), the output voltage of Plate-4 at 67.93 kHz increases with rising load resistance and eventually tends to saturate. Its saturation voltage is 0.914 V, while the saturation voltage of the reference pure-aluminum plate under identical settings is 0.302 V at 66.95 kHz. According to Ohm’s law, the maximum output powers of Phononic-Crystal Plate-4 and the pure-aluminum plate are calculated as 0.333 mW and 0.033 mW, respectively. The comparison shows that the voltage and power of Phononic-Crystal Plate-4 are approximately 3.03 times and 10.09 times those of the pure-aluminum plate.

Subsequently, the saturation voltage of Phononic-Crystal Plate-4 is compared with those of Plate-1, Plate-2 and Plate-3. It is found that the saturation voltage of Plate-4 is not only higher than those of Plate-1 and Plate-2 individually, but also approximately equal to the sum of their respective values. This observation verifies that Phononic-Crystal Plate-4 with large-area waveguides still retains the transmission characteristic of Phononic-Crystal Plate-3, i.e., the simultaneous propagation of both in-plane and out-of-plane elastic-wave components. Nevertheless, the saturation voltage of Plate-4 is slightly lower than that of Plate-3, reaching approximately 79.41 % of the latter. This originates from inevitable energy dissipation occurring with the enlargement of waveguide area, together with the limited size of the piezoelectric energy harvester at the receiver, which cannot capture energy over the full waveguide region and thus results in a slight drop in output performance.

Next, elastic-wave energy is harvested from Phononic-Crystal Plate-5 with a larger-area waveguide and Phononic-Crystal Plate-6 featuring gradient-varying large-area waveguides. Since the waveguide at the excitation side consists of three rows of Unit-Cell Type-A, three groups of excitation sources are introduced to excite along both the x- and z-directions. Nevertheless, the receiver and peripheral boundaries remain consistent with those of other phononic-crystal plates. As shown in Figure 6(e) and Figure 6(f), the output voltages of Phononic-Crystal Plate-5 at 66.83 kHz and Plate-6 at 66.23 kHz rise with increasing load resistance and finally reach saturation. Their saturation voltages are 1.321 V and 2.52 V, respectively. According to Ohm's law, the corresponding maximum output powers are calculated to be 0.916 mW and 1.914 mW. Compared with the saturation voltage of 0.302 V and maximum output power of 0.033 mW for the pure-aluminum plate, the voltage and power of Phononic-Crystal Plate-5 are approximately 4.37 times and 27.78 times those of the pure-aluminum plate, while those of Phononic-Crystal Plate-6 are approximately 8.34 times and 58 times those of the pure-aluminum plate. The experimental results demonstrate that adopting gradient variation in large-area waveguides can further realize energy convergence and enrichment, and greatly improve the energy-transmission volume and energy-harvesting efficiency.

## V. Conclusion

In this paper, a fully-polarized elastic topological heterostructure based on the quantum valley-Hall effect is proposed. It effectively addresses the technical challenges faced by conventional acoustic-energy-harvesting devices, including low harvesting efficiency and poor transmission stability, as well as the limitations of existing topological metamaterials that can only transmit the out-of-plane component of elastic waves, i.e., restricted large-volume energy transmission and severe energy dissipation.

The simultaneous and efficient full-polarization transmission of in-plane and out-of-plane components for vector elastic waves and high-efficiency energy harvesting in large-area waveguides are realized.

First, the degeneracy and breaking of the double Dirac cones for in-plane and out-of-plane modes are successfully realized by tuning the geometric parameters of unit cells. Multiple supercells are designed and constructed based on unit cells with different topological properties. The energy-harvesting performance of the structure with full-polarization transmission for both in-plane and out-of-plane components is remarkably superior to that of the structure supporting only out-of-plane components, which verifies the synergistic-enhancement effect originating from the coupled full-polarization transmission of elastic waves. Subsequently, unit cells with double Dirac cones are introduced at the topological boundaries to construct large-area waveguide states capable of transmitting full-polarization elastic-wave components. This greatly enlarges the energy-transmission volume while maintaining favorable transmission efficiency. Furthermore, the proposed gradient valley-locked structure achieves step-wise convergence and enrichment of energy for full-polarization elastic-wave components inside the waveguide. Its energy-harvesting performance reaches 4.79 times that of the conventional narrow-path waveguide structure, effectively solving the problem of low energy-harvesting efficiency under large-volume energy transmission. The proposed full-polarization elastic topological heterostructure enriches the wave-control theory and structural-design methodology of elastic topological metamaterials. It provides a new technical route for self-powered low-power devices in fields such as industrial-equipment monitoring and medical diagnosis and treatment.

**Appendix A. Energy-Transmission Measurement Experiments**

The experimental system consists of an Arbitrary-function generator, (DG1022U, RIGOL), a power amplifier (ATA-3080), CNC-machined aluminum-plate acoustic specimens, PZT piezoelectric ceramic sheets, and a three-dimensional scanning laser vibrometer (LV-FSC500-3D).

During the experiment, the Arbitrary-function generator outputs swept-frequency signals ranging from 66 kHz to 72 kHz. One channel is fed into the power amplifier, while the other channel is sent synchronously to the control and data-acquisition system as a reference signal. The amplified signal from the power amplifier drives the PZT piezoelectric-ceramic transducer to excite acoustic vibration modes in the acoustic specimen. The vibration-component signals of the specimen in three in-plane and out-of-plane directions are captured by the three-dimensional scanning laser vibrometer based on the laser-Doppler effect and then transmitted to the control and data-acquisition system. Finally, the amplitude and phase at each measuring point are calculated and analyzed from the two sets of signals, so as to investigate the acoustic characteristics of the waveguide.

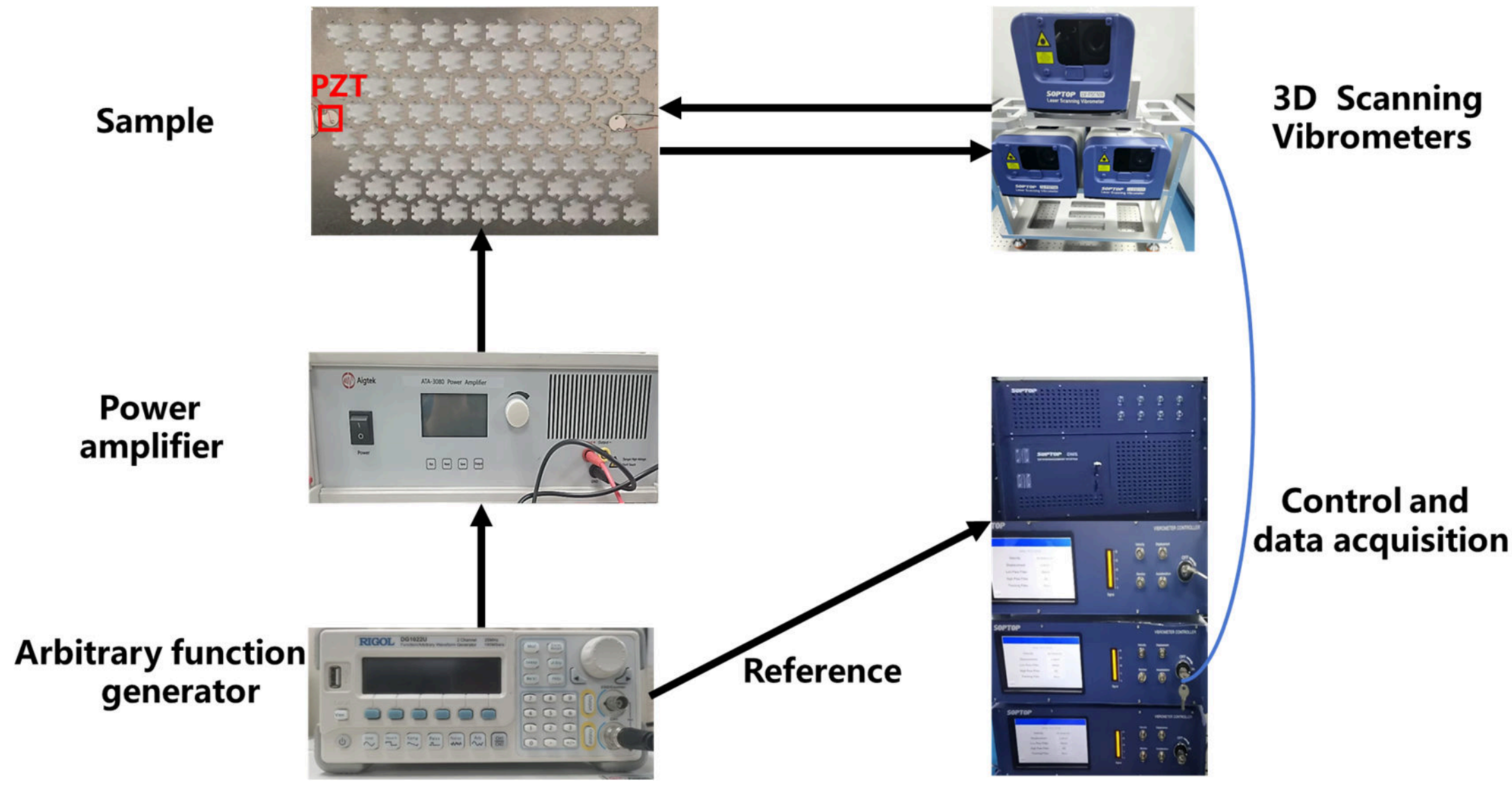

**FIG. A1.** Experimental setup

## Appendix B. Energy-Harvesting Measurement Experiments

The experimental system consists of an Aarbitrary-function generator, (DG1022U, RIGOL), a power amplifier (ATA-3080), CNC-machined aluminum-plate acoustic specimens, PZT piezoelectric ceramic sheets, a resistance box, and an oscilloscope (TDS1001C-SC).

During the experiment, the Aarbitrary-function generator first outputs swept-frequency signals ranging from 66 kHz to 72 kHz. After being amplified by the power amplifier, the signals drive the PZT piezoelectric-ceramic transducer at the input end of the acoustic specimen to convert electrical signals into mechanical vibrations.

Subsequently, the PZT piezoelectric-ceramic transducer at the output end converts the collected vibration signals of in-plane and out-of-plane full-polarization components into electrical signals applied to the resistance box. The oscilloscope measures the open-circuit voltage corresponding to each frequency, so as to obtain the optimal frequency with relatively high output voltage. Finally, the output frequency of the Aarbitrary-function generator is kept unchanged, and the relationship between the output voltage of the acoustic specimen and load resistance is investigated by adjusting the resistance value of the load resistor.

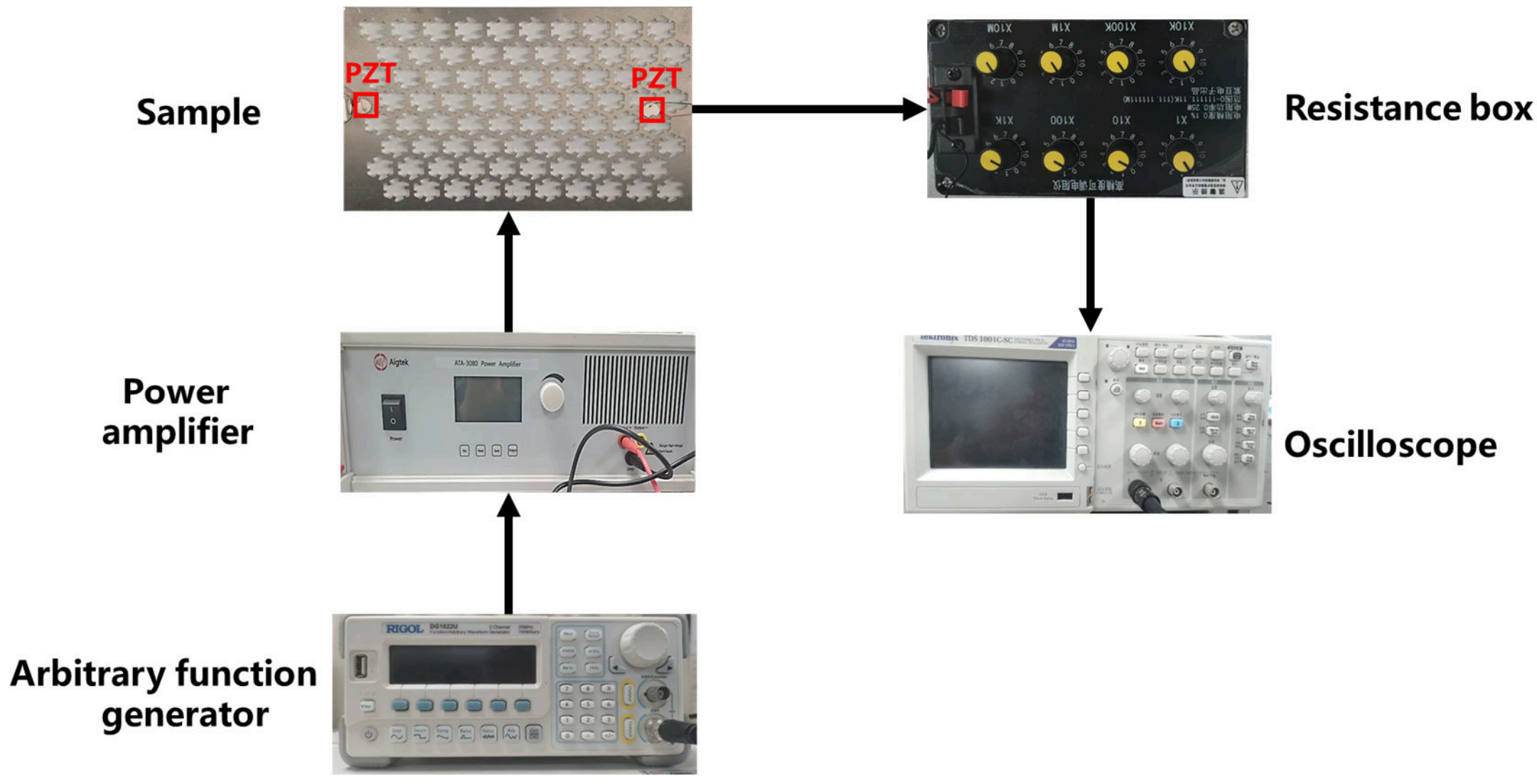

**FIG. B1.** Experimental setup